\documentclass[pra, aps, twocolumn, amsmath, amssymb, showpacs, superscriptaddress]{revtex4}

\usepackage{graphicx}
\usepackage{color}
\usepackage{bm}
\usepackage{graphics}

\begin{document}
\title{Hyperfine structure of alkali-metal diatomic molecules}

\author{Jesus Aldegunde}
\affiliation{Departamento de Quimica Fisica, Universidad de Salamanca, 37008
Salamanca, Spain}

\author{Jeremy M. Hutson}
\email{J.M.Hutson@durham.ac.uk} \affiliation{Joint Quantum Centre (JQC)
Durham-Newcastle, Department of Chemistry, Durham University, South Road,
Durham, DH1 3LE, United Kingdom}

\begin{abstract}
We present calculations of the hyperfine coupling constants for all the
heteronuclear alkali-metal diatomic molecules at the equilibrium geometry of
the electronic ground state. These constants are important in developing
methods to control ultracold polar molecules. The results are based on
electronic structure calculations using density functional theory, and are in
good agreement with experiment for the limited set of molecules for which
experiments are so far available.
\end{abstract}

\date{\today}

\maketitle

\section{Introduction}

It has recently become possible to produce samples of ultracold polar molecules
at temperatures around 1 $\mu$K, by combining pairs of ultracold alkali-metal
atoms by magnetoassociation and then transferring them to their rovibronic
ground state by Stimulated Raman Adiabatic Passage (STIRAP). The polar
molecules produced so far include $^{40}$K$^{87}$Rb \cite{Ni:KRb:2008},
$^{87}$Rb$^{133}$Cs \cite{Takekoshi:RbCs:2014, Molony:RbCs:2014},
$^{23}$Na$^{40}$K \cite{Park:NaK:2015} and $^{23}$Na$^{87}$Rb
\cite{Guo:NaRb:2016}. Such molecules have many potential applications, ranging
from quantum-state-controlled chemistry \cite{Krems:PCCP:2008,
Ospelkaus:hyperfine-control:2010, Ni:2010, deMiranda:2011}, to quantum
simulation \cite{Santos:2000, Baranov:2012} and quantum information
\cite{DeMille:2002,Yelin:2006}.

All the stable isotopes of the alkali metals have non-zero nuclear spin. In the
diatomic molecules, the two spins interact with one another and with the
molecular rotation to form complex patterns of energy levels. These energy
levels cross and avoided-cross as a function of magnetic and electric fields
\cite{Aldegunde:polar:2008, Aldegunde:spectra:2009, Ran:2010} and laser
intensity \cite{Gregory:RbCs-AC-Stark:2017}. Understanding the energy levels
and their crossings is crucial in developing schemes to control ultracold
molecules and transfer them between rotational and hyperfine states.

We have previously carried out calculations of the hyperfine coupling constants
of KRb and RbCs \cite{Aldegunde:polar:2008} and LiCs \cite{Ran:2010}, using
density-functional theory. The purpose of the present paper is to extend these
calculations to the full set of heteronuclear diatomic molecules formed from
alkali-metal atoms. We compare the results with experiments
\cite{Ospelkaus:hyperfine-control:2010, Will:2016, Gregory:RbCs-microwave:2016}
where possible.

\section{Molecular Hamiltonian} \label{sec:MolHam}

The effective Hamiltonian of a $^1\Sigma$ diatomic molecule, with hyperfine
structure, in the presence of external magnetic and electric fields may be
written \cite{Ramsey:1952, Brown:2003, Bryce:2003, Aldegunde:polar:2008,
Gregory:RbCs-AC-Stark:2017}
\begin{equation}
\label{Htot} H = H_{\rm rot} + H_{\rm hf}  + H_{\rm S}  + H_{\rm Z},
\end{equation}
where $H_{\rm rot}$, $H_{\rm hf}$, $H_{\rm S}$ and $H_{\rm Z}$ are rotational,
hyperfine, Stark and Zeeman terms,
\begin{eqnarray}
 \label{eq:Hrot} H_{\rm rot}  &=& B_v\bm{N}^{2}-D_v\bm{N}^{2}\cdot\bm{N}^{2}; \\
  \nonumber \label{eq:Hhf}H_{\rm hf}  &=& \sum_{i=1}^{2}\bm{V}_{i}:\bm{Q}_{i} \\
&+& \sum_{i=1}^{2} c_{i} \,\bm{N}\cdot\bm{I}_{i}
+c_{3}\,\bm{I}_{1}\cdot\bm{T}\cdot\bm{I}_{2}
+c_{4}\,\bm{I}_{1}\cdot\bm{I}_{2};\\
\label{eq:Hs} H_{\rm S}  &=& -\bm{\mu}\cdot\bm{E}
- \frac{1}{2}\bm{E}\cdot\bm{\alpha}\cdot\bm{E};\\
\label{eq:Hz} H_{\rm Z}  &=& -g_{\rm r}\mu_{\rm N}
\,\bm{N}\cdot\bm{B}
 -\sum_{i=1}^{2}g_{i}\mu_{\rm N}
\,\bm{I}_{i}\cdot\bm{B} (1-\sigma_{i}).
\end{eqnarray}
Here $\bm{I}_{1}$ and $\bm{I}_{2}$ are the spins of nuclei 1 and 2 and $\bm{N}$
is the angular momentum for rotation of the
molecule about its center of mass. The rotational and centrifugal distortion
constants of the molecule are $B_v$ and $D_v$, though centrifugal distortion is
neglected below. The hyperfine Hamiltonian \eqref{eq:Hhf} consists of four
terms. The first is the interaction between the nuclear electric quadrupole
tensor $\bm{Q}$ and the electric field gradient $\bm{V}$ due to the electrons,
which is characterized by coupling constants $(eqQ)_{1}$ and $(eqQ)_{2}$. The
second is the interaction between the nuclear magnetic moments and the magnetic
field created by the rotation of the molecule, with spin-rotation coupling
constants $c_{1}$ and $c_{2}$. The final two terms are the tensor and scalar
interactions between the nuclear dipole moments, with spin-spin coupling
constants $c_{3}$ and $c_{4}$ respectively.

The Stark Hamiltonian (\ref{eq:Hs}) includes both a linear term to describe the
interaction of the molecular dipole $\bm{\mu}$ with a static electric field
$\bm{E}$ and a quadratic term involving the molecular polarizability tensor
$\bm{\alpha}$. The latter is usually small for static fields, but may be used
with a frequency-dependent polarizability $\bm{\alpha}(\omega)$ to account for
the ac Stark effect due to a non-resonant laser field
\cite{Gregory:RbCs-AC-Stark:2017}. The Zeeman Hamiltonian (\ref{eq:Hz})
describes the interaction of the molecule with an external magnetic field
$\bm{B}$, and consists of two terms representing the rotational and nuclear
Zeeman effects. In the latter, $g_{i}$ and $\sigma_i$ are the nuclear g-factor
and the shielding factor for nucleus $i$.

\section{Evaluation of the coupling constants}
\label{sec:eval}

In the present work we evaluate the hyperfine coupling constants from
electronic structure calculations using density-functional theory (DFT). The
methods used are the same as in ref.\ \cite{Aldegunde:polar:2008}, so will be
summarized only briefly here. The calculations are performed with the Amsterdam
Density Functional (ADF) package \cite{ADF1,ADF3}. We employ all-electron QZ4P
basis sets (quadruple-$\zeta$ basis sets with four polarization functions).
Relativistic corrections are included by means of the two-component zero-order
regular approximation (ZORA) \cite{vanLenthe:1993, vanLenthe:1994,
vanLenthe:1999}, including spin-orbit coupling as well as scalar effects. We
use different density functionals for different properties: quadrupole coupling
constants are obtained from calculations with the B3LYP functional
\cite{lee:1988, becke:1993}, while spin-spin coupling constants are obtained
with the PBE functional \cite{perdew:1996}. Shielding tensors and the related
spin-rotation constants are evaluated using the KT2 functional
\cite{Keal:2003}. All these choices are justified in ref.\
\cite{Aldegunde:polar:2008}.

ADF generally calculates hyperfine coupling constants for the most common
isotope of each element. In the present work we provide values for all
combinations of stable isotopes by performing simple scalings according to
nuclear g-factors, nuclear quadrupole moments \cite{Mills:1988, Stone:2016} and
molecular rotational constants. The results obtained are given in Table
\ref{tb:dimers} \footnote{Values for $^7$Li$^{133}$Cs, KRb and RbCs, obtained
with similar methods, were presented previously \cite{Aldegunde:polar:2008,
Ran:2010}. The values in Table \ref{tb:dimers} differ slightly (by around 1\%
except for some values of $eQq$) because we used tighter DFT convergence
criteria and updated nuclear quadrupoles \cite{Stone:2016} in the present
work.}. The calculations were performed at the equilibrium geometry for each
molecule, $R_{\rm{e}}$=2.88 \AA\ for LiNa \cite{Engelke:1982}, 3.32 \AA\ for
LiK \cite{Martin:2001}, 3.43 \AA\ for LiRb \cite{Korek:2000}, 3.67 \AA\ for
LiCs \cite{Staanum:2007}, 3.45 \AA\ for NaK \cite{Wormsbecher:1981}, 3.64 \AA\
for NaRb \cite{Docenko:NaRb:2004}, 3.85 \AA\ for NaCs \cite{Docenko:2006}, 4.07
\AA\ for KRb \cite{Ross:1990}, 4.28 \AA\ for KCs \cite{Ferber:2008} and 4.37
\AA\ for RbCs \cite{Kato:1983}. The values of the permanent dipole moments, not
included in Table \ref{tb:dimers}, can be found in Ref.\ \cite{Fedorov:2014}.

\section{Comparison with experiment}
\label{sec:hspli}

Experimental determinations of the hyperfine coupling constants
from molecular spectroscopy are mostly limited to the scalar spin-spin coupling
constant $c_4$ and the nuclear quadrupole coupling constants $(eQq)_1$ and
$(eQq)_2$. In particular, the ground rotational state ($N=0$) is almost
unaffected by any hyperfine couplings except $c_4$ \cite{Aldegunde:polar:2008}.
At zero magnetic field it splits into $2I_{\rm min}+1$ states, where $I_{\rm
min}$ is the smaller of $I_1$ and $I_2$. These correspond to the different
possible values of the total angular momentum $F$, which for $N=0$ is the same
as the total nuclear spin $I$. The splitting between the highest and lowest
states due to the hyperfine coupling is
\begin{equation}\label{eq:N0sp}
  \Delta E_{N=0}^{\rm hf}=\frac{|c_4|}{2}[(I_1+I_2)(I_1+I_2+1)-|I_1-I_2|(|I_1-I_2|+1)]
\end{equation}
and ranges from 1.3 kHz for $^6$Li$^{41}$K to 208 kHz for $^{87}$Rb$^{133}$Cs.

For $N>0$ the hyperfine splitting is more complicated and is commonly dominated
by the nuclear electric quadrupole interaction, with a minor contribution from
the scalar spin-spin interaction. Under these circumstances, the splitting
between the highest and lowest $N=1$ levels, $\Delta E_{N=1}^{\rm hf}$, is
approximately half the larger of the two nuclear electric quadrupole coupling
constants. This can be as much as 4 MHz for $^{6}$Li$^{85}$Rb and
$^{7}$Li$^{85}$Rb. For some molecules such as $^{23}$Na$^{133}$Cs or
$^{7}$Li$^{133}$Cs, however, the nuclear electric quadrupole interactions are
small enough (tens of kHz) to be comparable to the scalar spin-spin
interaction, and the two effects influence the hyperfine splittings by similar
amounts. The tensorial spin-spin and spin-rotation interactions play a minor
role for low-$N$ states, though the spin-rotation interaction may become
significant for experiments involving higher rotational levels.

Table \ref{tb:expvsth} compares the calculated hyperfine constants to
experimental values where available. The calculated values are mostly within
10\% of experiment, and always within 20\%, except for $(eQq)_{\rm Na}$ in
$^{23}$Na$^{40}$K and $(eQq)_{\rm K}$ in $^{40}$K$^{87}$Rb, which are
experimentally less well determined than most other constants.

\section{Conclusions}
\label{sec:conc}

We have presented calculations of the hyperfine coupling constants for all
heteronuclear alkali-metal diatomic molecules formed from stable isotopes,
using electronic structure calculations based on density functional theory.
Characterizing the hyperfine structure of these molecules is essential to
controlling them and developing their applications in ultracold quantum
physics. Our results are in good agreement with the (still scarce) experimental
measurements of these molecular properties.

\begin{table*}
\caption{Nuclear properties and rotational and hyperfine coupling
constants for the heteronuclear alkali-metal diatomic molecules. Indices 1 and
2 refer to the first and second atom respectively. Calculations were performed
as described in the text.}
\label{tb:dimers}       
\begin{tabular}{ cccccccccccccc }
\hline\noalign{\smallskip}
 & $I_{1}$ & $I_{2}$ & $g_{1}$ & $g_{2}$ & $B_v$(GHz) & $(eQq)_{1}$(MHz)
 & $(eQq)_{2}$(MHz) & $\sigma_{1}$(ppm) & $\sigma_{2}$(ppm) & $c_{1}$(Hz)
 & $c_{2}$(Hz) & $c_{3}$(Hz) & $c_{4}$(Hz) \\
\noalign{\smallskip}\hline\noalign{\smallskip}
$^{6}\rm{Li}^{23}\rm{Na}$ & 1 & 3/2 & 0.822 & 1.478 & 12.735 & $8\times10^{-4}$ & $-$0.684 & 102.3 & 613.6 &
84.0  & 805.3 & 196.6 & 212.2 \\ \noalign{\smallskip}
$^{7}\rm{Li}^{23}\rm{Na}$ & 3/2 & 3/2 & 2.171 & 1.478 & 11.296 & 0.038 & $-$0.684 & 102.3
 & 613.6 & 196.9 & 714.2 & 519.2 & 560.4  \\ \noalign{\smallskip}
$^{6}\rm{Li}^{39}\rm{K}$ & 1 & 3/2 & 0.822 & 0.261 & 8.799 & $4\times10^{-4}$ & $-$0.854 & 104.1
 & 1296.8 & 48.3 & 238.8 & 23.1 & 58.7  \\ \noalign{\smallskip}
$^{6}\rm{Li}^{40}\rm{K}$ & 1 & 4 & 0.822 & $-$0.324 & 8.770 & $4\times10^{-4}$ & 1.066 & 104.1 & 1296.8
 & 48.2 & $-$295.5 & $-$28.7 & $-$72.8 \\ \noalign{\smallskip}
$^{6}\rm{Li}^{41}\rm{K}$ & 1 & 3/2 & 0.822 & 0.143 & 8.742 & $4\times10^{-4}$ & $-$1.038 & 104.1 & 1296.8
 & 48.0 & 130.0 & 12.6 & 32.1 \\ \noalign{\smallskip}
$^{7}\rm{Li}^{39}\rm{K}$ & 3/2 & 3/2 & 2.171 & 0.261 & 7.712 & 0.021 & $-$0.854 & 104.1 & 1296.8
 & 111.9 & 209.3 & 61.0 & 155.0 \\ \noalign{\smallskip}
$^{7}\rm{Li}^{40}\rm{K}$ & 3/2 & 4 & 2.171 & $-$0.324 & 7.682 & 0.021 & 1.066 & 104.1 & 1296.8
 & 111.5 & $-$258.9 & $-$75.7 & $-$192.4 \\ \noalign{\smallskip}
$^{7}\rm{Li}^{41}\rm{K}$ & 3/2 & 3/2 & 2.171 & 0.143 & 7.654 & 0.021 & $-$1.038 & 104.1 & 1296.8
 & 111.1 & 113.8 & 33.4 & 84.9 \\ \noalign{\smallskip}
$^{6}\rm{Li}^{85}\rm{Rb}$ & 1 & 5/2 & 0.822 & 0.541 & 7.647 & $2\times10^{-4}$ & $-$7.774 & 105.4 & 3419.8
 & 36.0 & 1177.0 & 44.7 & 261.4 \\ \noalign{\smallskip}
$^{6}\rm{Li}^{87}\rm{Rb}$ & 1 & 3/2 & 0.822 & 1.834 & 7.636 & $2\times10^{-4}$ & $-$3.760 & 105.4 & 3419.8
 & 36.0 & 3983.8 & 151.7 & 886.3 \\ \noalign{\smallskip}
$^{7}\rm{Li}^{85}\rm{Rb}$ & 3/2 & 5/2 & 2.171 & 0.541 & 6.628 & 0.012 & $-$7.774 & 105.4 & 3419.8
 & 82.5 & 1020.2 & 118.1 & 690.5 \\ \noalign{\smallskip}
$^{7}\rm{Li}^{87}\rm{Rb}$ & 3/2 & 3/2 & 2.171 & 1.834 & 6.617 & 0.012 & $-$3.760 & 105.4 & 3419.8
 & 82.3 & 3452.3 & 400.7 & 2340.9 \\ \noalign{\smallskip}
$^{6}\rm{Li}^{133}\rm{Cs}$ & 1   & 7/2 & 0.822 & 0.738 & 6.520 & $3\times10^{-4}$ & 0.181 & 108.0 & 6244.0
 & 15.2 & 3475.5 & 53.1  & 620.8  \\ \noalign{\smallskip}
$^{7}\rm{Li}^{133}\rm{Cs}$ & 3/2 & 7/2 & 2.171 & 0.738 & 5.630 &   0.017          & 0.181 & 108.0 & 6244.0
 & 34.7 & 3001.2 & 140.1 & 1639.7 \\ \noalign{\smallskip}
$^{23}\rm{Na}^{39}\rm{K}$ & 3/2 & 3/2 & 1.478 & 0.261 & 2.937 & $-$0.133 & $-$0.613 & 624.4 & 1297.4
 & 118.5 & 78.9  & 39.0  & 362.5  \\ \noalign{\smallskip}
$^{23}\rm{Na}^{40}\rm{K}$ & 3/2 & 4   & 1.478 & $-$0.324& 2.909 & $-$0.133 & 0.765  & 624.4 & 1297.4
 & 117.4 & $-$97.0 & $-$48.4 & $-$450.0  \\ \noalign{\smallskip}
$^{23}\rm{Na}^{41}\rm{K}$ & 3/2 & 3/2 & 1.478 & 0.143 & 2.883 & $-$0.133 & $-$0.745 & 624.4 & 1297.4
 & 116.4 & 42.4  & 21.4  & 198.6  \\ \noalign{\smallskip}
$^{23}\rm{Na}^{85}\rm{Rb}$ & 3/2 & 5/2 & 1.478 & 0.541 & 2.108 & $-$0.132 & $-$6.170 & 629.6 & 3437.6
 & 61.0 & 291.6 & 76.5 & 1690.3  \\ \noalign{\smallskip}
$^{23}\rm{Na}^{87}\rm{Rb}$ & 3/2 & 3/2 & 1.478 & 1.834 & 2.098 & $-$0.132 & $-$2.984 & 629.6 & 3437.6
 & 60.7 & 983.8 & 259.3 & 5730.3  \\ \noalign{\smallskip}
$^{23}\rm{Na}^{133}\rm{Cs}$ & 3/2& 7/2 & 1.478 & 0.738 & 1.740 & $-$0.097 & 0.150  & 639.2 & 6278.7
 & 14.2 & 854.5 & 105.6& 3941.8  \\ \noalign{\smallskip}
$^{39}\rm{K}^{85}\rm{Rb}$ & 3/2 & 5/2 & 0.261 & 0.541 & 1.142 & $-$0.249 & $-$3.066 & 1321.0 & 3469.0
 & 19.9  & 126.8 & 11.5 & 482.1   \\ \noalign{\smallskip}
$^{39}\rm{K}^{87}\rm{Rb}$ & 3/2 & 3/2 & 0.261 & 1.834 & 1.134 & $-$0.249 & $-$1.483 & 1321.0 & 3469.0
 & 19.8  & 426.9 & 38.8 & 1634.3  \\ \noalign{\smallskip}
$^{40}\rm{K}^{85}\rm{Rb}$ & 4   & 5/2 &-0.324 & 0.541 & 1.123 & 0.311  & $-$3.066 & 1321.0 & 3469.0
 & $-$24.3 & 124.7 & $-$14.2& $-$598.5 \\ \noalign{\smallskip}
$^{40}\rm{K}^{87}\rm{Rb}$ & 4   & 3/2 &-0.324 & 1.834 & 1.114 & 0.311  & $-$1.483 & 1321.0 & 3469.0
 & $-$24.1 & 419.5 & $-$48.2& $-$2028.8 \\ \noalign{\smallskip}
$^{41}\rm{K}^{85}\rm{Rb}$ & 3/2 & 5/2 & 0.143 & 0.541 & 1.104 & $-$0.303 & $-$3.066 & 1321.0 & 3469.0
 & 10.5  & 122.6 & 6.3  & 264.1  \\ \noalign{\smallskip}
$^{41}\rm{K}^{87}\rm{Rb}$ & 3/2 & 3/2 & 0.143 & 1.834 & 1.096 & $-$0.303 & $-$1.483 & 1321.0 & 3469.0
 & 10.5  & 412.5 & 21.3 & 895.4  \\ \noalign{\smallskip}
$^{39}\rm{K}^{133}\rm{Cs}$ & 3/2 & 7/2 & 0.261 & 0.738 & 0.916 & $-$0.182 & 0.075 & 1340.7 & 6337.1
 & 8.6  & 385.4 & 18.0 & 1146.3 \\ \noalign{\smallskip}
$^{40}\rm{K}^{133}\rm{Cs}$ & 4   & 7/2 & $-$0.324& 0.738 & 0.898 &  0.227 & 0.075 & 1340.7 & 6337.1
 & $-$10.5& 377.9 & $-$22.3& $-$1422.9\\ \noalign{\smallskip}
$^{41}\rm{K}^{133}\rm{Cs}$ & 3/2 & 7/2 & 0.143 & 0.738 & 0.881 & $-$0.221 & 0.075 & 1340.7 & 6337.1
 & 4.5  & 370.8 & 9.9  & 628.0  \\ \noalign{\smallskip}
$^{85}\rm{Rb}^{133}\rm{Cs}$ & 5/2 & 7/2 & 0.541 & 0.738 & 0.511 & $-$1.611 & 0.054 & 3531.6 & 6367.3
 & 29.2 & 196.5 & 56.8 & 5116.7 \\ \noalign{\smallskip}
$^{87}\rm{Rb}^{133}\rm{Cs}$ & 3/2 & 7/2 & 1.834 & 0.738 & 0.504 & $-$0.779 & 0.054 & 3531.6 & 6367.3
 & 97.6 & 193.7 & 192.5 & 17345.8 \\ \noalign{\smallskip}
\noalign{\smallskip}\hline
\end{tabular}
\end{table*}

\begin{table*}
\caption{Comparison between experimental and theoretical values of molecular
constants. Indices 1 and 2 refer to the first and second atom respectively.}
\label{tb:expvsth}
\begin{tabular}{ cccccc }
\hline\noalign{\smallskip}
 & $B_v$(GHz) & $(eQq)_{1}$(MHz) & $(eQq)_{2}$(MHz) & $c_{4}$(Hz) & Reference \\
\noalign{\smallskip}\hline\noalign{\smallskip}
$^{40}$K$^{87}$Rb &1.113950(5)&0.45(6)&$-$1.41(4)&---& \cite{Ospelkaus:hyperfine-control:2010} \\
 &1.114&0.311&$-$1.483&---& This work and \cite{Aldegunde:polar:2008}\\
 \hline
$^{23}$Na$^{40}$K &2.8217297(10)&$-$0.187(35)&0.899(20)&$-$409(10)& \cite{Will:2016} \\
 &2.909&$-$0.133&0.765&$-$450.0& This work \\
 \hline
$^{87}$Rb$^{133}$Cs &0.490173994(45)&$-$0.80929(113)&0.05998(186)&19019(105)& \cite{Gregory:RbCs-microwave:2016} \\
 &0.504&$-$0.779&0.054&17345.8& This work and \cite{Aldegunde:polar:2008} \\
 \hline
\end{tabular}
\end{table*}

\begin{acknowledgments}
This work was supported by the U.K. Engineering and Physical Sciences Research
Council (EPSRC) Grants No.\ EP/H003363/1, EP/I012044/1, EP/P008275/1 and
EP/P01058X/1. JA acknowledges funding by the Spanish Ministry of Science and
Innovation Grants No.\ CTQ2012-37404-C02, CTQ2015-65033-P, and Consolider
Ingenio 2010 CSD2009-00038.
\end{acknowledgments}



\begin{thebibliography}{44}
\expandafter\ifx\csname natexlab\endcsname\relax\def\natexlab#1{#1}\fi
\expandafter\ifx\csname bibnamefont\endcsname\relax
  \def\bibnamefont#1{#1}\fi
\expandafter\ifx\csname bibfnamefont\endcsname\relax
  \def\bibfnamefont#1{#1}\fi
\expandafter\ifx\csname citenamefont\endcsname\relax
  \def\citenamefont#1{#1}\fi
\expandafter\ifx\csname url\endcsname\relax
  \def\url#1{\texttt{#1}}\fi
\expandafter\ifx\csname urlprefix\endcsname\relax\def\urlprefix{URL }\fi
\providecommand{\bibinfo}[2]{#2}
\providecommand{\eprint}[2][]{\url{#2}}

\bibitem[{\citenamefont{Ni et~al.}(2008)\citenamefont{Ni, Ospelkaus, {de
  Miranda}, Pe'er, Neyenhuis, Zirbel, Kotochigova, Julienne, Jin, and
  Ye}}]{Ni:KRb:2008}
\bibinfo{author}{\bibfnamefont{K.-K.} \bibnamefont{Ni}},
  \bibinfo{author}{\bibfnamefont{S.}~\bibnamefont{Ospelkaus}},
  \bibinfo{author}{\bibfnamefont{M.~H.~G.} \bibnamefont{{de Miranda}}},
  \bibinfo{author}{\bibfnamefont{A.}~\bibnamefont{Pe'er}},
  \bibinfo{author}{\bibfnamefont{B.}~\bibnamefont{Neyenhuis}},
  \bibinfo{author}{\bibfnamefont{J.~J.} \bibnamefont{Zirbel}},
  \bibinfo{author}{\bibfnamefont{S.}~\bibnamefont{Kotochigova}},
  \bibinfo{author}{\bibfnamefont{P.~S.} \bibnamefont{Julienne}},
  \bibinfo{author}{\bibfnamefont{D.~S.} \bibnamefont{Jin}}, \bibnamefont{and}
  \bibinfo{author}{\bibfnamefont{J.}~\bibnamefont{Ye}},
  \bibinfo{journal}{Science} \textbf{\bibinfo{volume}{322}},
  \bibinfo{pages}{231} (\bibinfo{year}{2008}).

\bibitem[{\citenamefont{Takekoshi et~al.}(2014)\citenamefont{Takekoshi,
  Reichs\"ollner, Schindewolf, Hutson, {Le Sueur}, Dulieu, Ferlaino, Grimm, and
  N\"agerl}}]{Takekoshi:RbCs:2014}
\bibinfo{author}{\bibfnamefont{T.}~\bibnamefont{Takekoshi}},
  \bibinfo{author}{\bibfnamefont{L.}~\bibnamefont{Reichs\"ollner}},
  \bibinfo{author}{\bibfnamefont{A.}~\bibnamefont{Schindewolf}},
  \bibinfo{author}{\bibfnamefont{J.~M.} \bibnamefont{Hutson}},
  \bibinfo{author}{\bibfnamefont{C.~R.} \bibnamefont{{Le Sueur}}},
  \bibinfo{author}{\bibfnamefont{O.}~\bibnamefont{Dulieu}},
  \bibinfo{author}{\bibfnamefont{F.}~\bibnamefont{Ferlaino}},
  \bibinfo{author}{\bibfnamefont{R.}~\bibnamefont{Grimm}}, \bibnamefont{and}
  \bibinfo{author}{\bibfnamefont{H.-C.} \bibnamefont{N\"agerl}},
  \bibinfo{journal}{Phys. Rev. Lett.} \textbf{\bibinfo{volume}{113}},
  \bibinfo{pages}{205301} (\bibinfo{year}{2014}).

\bibitem[{\citenamefont{Molony et~al.}(2014)\citenamefont{Molony, Gregory, Ji,
  Lu, K\"oppinger, {Le Sueur}, Blackley, Hutson, and
  Cornish}}]{Molony:RbCs:2014}
\bibinfo{author}{\bibfnamefont{P.~K.} \bibnamefont{Molony}},
  \bibinfo{author}{\bibfnamefont{P.~D.} \bibnamefont{Gregory}},
  \bibinfo{author}{\bibfnamefont{Z.}~\bibnamefont{Ji}},
  \bibinfo{author}{\bibfnamefont{B.}~\bibnamefont{Lu}},
  \bibinfo{author}{\bibfnamefont{M.~P.} \bibnamefont{K\"oppinger}},
  \bibinfo{author}{\bibfnamefont{C.~R.} \bibnamefont{{Le Sueur}}},
  \bibinfo{author}{\bibfnamefont{C.~L.} \bibnamefont{Blackley}},
  \bibinfo{author}{\bibfnamefont{J.~M.} \bibnamefont{Hutson}},
  \bibnamefont{and} \bibinfo{author}{\bibfnamefont{S.~L.}
  \bibnamefont{Cornish}}, \bibinfo{journal}{Phys. Rev. Lett.}
  \textbf{\bibinfo{volume}{113}}, \bibinfo{pages}{255301}
  (\bibinfo{year}{2014}).

\bibitem[{\citenamefont{Park et~al.}(2015)\citenamefont{Park, Will, and
  Zwierlein}}]{Park:NaK:2015}
\bibinfo{author}{\bibfnamefont{J.~W.} \bibnamefont{Park}},
  \bibinfo{author}{\bibfnamefont{S.~A.} \bibnamefont{Will}}, \bibnamefont{and}
  \bibinfo{author}{\bibfnamefont{M.~W.} \bibnamefont{Zwierlein}},
  \bibinfo{journal}{Phys. Rev. Lett.} \textbf{\bibinfo{volume}{114}},
  \bibinfo{pages}{205302} (\bibinfo{year}{2015}).

\bibitem[{\citenamefont{Guo et~al.}(2016)\citenamefont{Guo, Zhu, Lu, Ye, Wang,
  Vexiau, Bouloufa-Maafa, Qu\'em\'ener, Dulieu, and Wang}}]{Guo:NaRb:2016}
\bibinfo{author}{\bibfnamefont{M.}~\bibnamefont{Guo}},
  \bibinfo{author}{\bibfnamefont{B.}~\bibnamefont{Zhu}},
  \bibinfo{author}{\bibfnamefont{B.}~\bibnamefont{Lu}},
  \bibinfo{author}{\bibfnamefont{X.}~\bibnamefont{Ye}},
  \bibinfo{author}{\bibfnamefont{F.}~\bibnamefont{Wang}},
  \bibinfo{author}{\bibfnamefont{R.}~\bibnamefont{Vexiau}},
  \bibinfo{author}{\bibfnamefont{N.}~\bibnamefont{Bouloufa-Maafa}},
  \bibinfo{author}{\bibfnamefont{G.}~\bibnamefont{Qu\'em\'ener}},
  \bibinfo{author}{\bibfnamefont{O.}~\bibnamefont{Dulieu}}, \bibnamefont{and}
  \bibinfo{author}{\bibfnamefont{D.}~\bibnamefont{Wang}},
  \bibinfo{journal}{Phys. Rev. Lett.} \textbf{\bibinfo{volume}{116}},
  \bibinfo{pages}{205303} (\bibinfo{year}{2016}).

\bibitem[{\citenamefont{Krems}(2008)}]{Krems:PCCP:2008}
\bibinfo{author}{\bibfnamefont{R.~V.} \bibnamefont{Krems}},
  \bibinfo{journal}{Phys. Chem. Chem. Phys.} \textbf{\bibinfo{volume}{10}},
  \bibinfo{pages}{4079} (\bibinfo{year}{2008}).

\bibitem[{\citenamefont{Ospelkaus et~al.}(2010)\citenamefont{Ospelkaus, Ni,
  Qu\'{e}m\'{e}ner, Neyenhuis, Wang, {de Miranda}, Bohn, Ye, and
  Jin}}]{Ospelkaus:hyperfine-control:2010}
\bibinfo{author}{\bibfnamefont{S.}~\bibnamefont{Ospelkaus}},
  \bibinfo{author}{\bibfnamefont{K.-K.} \bibnamefont{Ni}},
  \bibinfo{author}{\bibfnamefont{G.}~\bibnamefont{Qu\'{e}m\'{e}ner}},
  \bibinfo{author}{\bibfnamefont{B.}~\bibnamefont{Neyenhuis}},
  \bibinfo{author}{\bibfnamefont{D.}~\bibnamefont{Wang}},
  \bibinfo{author}{\bibfnamefont{M.~H.~G.} \bibnamefont{{de Miranda}}},
  \bibinfo{author}{\bibfnamefont{J.~L.} \bibnamefont{Bohn}},
  \bibinfo{author}{\bibfnamefont{J.}~\bibnamefont{Ye}}, \bibnamefont{and}
  \bibinfo{author}{\bibfnamefont{D.~S.} \bibnamefont{Jin}},
  \bibinfo{journal}{Phys. Rev. Lett.} \textbf{\bibinfo{volume}{104}},
  \bibinfo{pages}{030402} (\bibinfo{year}{2010}).

\bibitem[{\citenamefont{Ni et~al.}(2010)\citenamefont{Ni, Ospelkaus, Wang,
  Qu\'em\'ener, Neyenhuis, {de Miranda}, Bohn, Ye, and Jin}}]{Ni:2010}
\bibinfo{author}{\bibfnamefont{K.-K.} \bibnamefont{Ni}},
  \bibinfo{author}{\bibfnamefont{S.}~\bibnamefont{Ospelkaus}},
  \bibinfo{author}{\bibfnamefont{D.}~\bibnamefont{Wang}},
  \bibinfo{author}{\bibfnamefont{G.}~\bibnamefont{Qu\'em\'ener}},
  \bibinfo{author}{\bibfnamefont{B.}~\bibnamefont{Neyenhuis}},
  \bibinfo{author}{\bibfnamefont{M.~H.~G.} \bibnamefont{{de Miranda}}},
  \bibinfo{author}{\bibfnamefont{J.~L.} \bibnamefont{Bohn}},
  \bibinfo{author}{\bibfnamefont{J.}~\bibnamefont{Ye}}, \bibnamefont{and}
  \bibinfo{author}{\bibfnamefont{D.~S.} \bibnamefont{Jin}},
  \bibinfo{journal}{Nature} \textbf{\bibinfo{volume}{464}},
  \bibinfo{pages}{1324} (\bibinfo{year}{2010}).

\bibitem[{\citenamefont{de~Miranda et~al.}(2011)\citenamefont{de~Miranda,
  Chotia, Neyenhuis, Wang, Qu\'em\'ener, Ospelkaus, Bohn, Ye, and
  Jin}}]{deMiranda:2011}
\bibinfo{author}{\bibfnamefont{M.~H.~G.} \bibnamefont{de~Miranda}},
  \bibinfo{author}{\bibfnamefont{A.}~\bibnamefont{Chotia}},
  \bibinfo{author}{\bibfnamefont{B.}~\bibnamefont{Neyenhuis}},
  \bibinfo{author}{\bibfnamefont{D.}~\bibnamefont{Wang}},
  \bibinfo{author}{\bibfnamefont{G.}~\bibnamefont{Qu\'em\'ener}},
  \bibinfo{author}{\bibfnamefont{S.}~\bibnamefont{Ospelkaus}},
  \bibinfo{author}{\bibfnamefont{J.~L.} \bibnamefont{Bohn}},
  \bibinfo{author}{\bibfnamefont{J.}~\bibnamefont{Ye}}, \bibnamefont{and}
  \bibinfo{author}{\bibfnamefont{D.~S.} \bibnamefont{Jin}},
  \bibinfo{journal}{Nat. Phys.} \textbf{\bibinfo{volume}{7}},
  \bibinfo{pages}{502} (\bibinfo{year}{2011}).

\bibitem[{\citenamefont{Santos et~al.}(2000)\citenamefont{Santos, Shlyapnikov,
  Zoller, and Lewenstein}}]{Santos:2000}
\bibinfo{author}{\bibfnamefont{L.}~\bibnamefont{Santos}},
  \bibinfo{author}{\bibfnamefont{G.~V.} \bibnamefont{Shlyapnikov}},
  \bibinfo{author}{\bibfnamefont{P.}~\bibnamefont{Zoller}}, \bibnamefont{and}
  \bibinfo{author}{\bibfnamefont{M.}~\bibnamefont{Lewenstein}},
  \bibinfo{journal}{Phys. Rev. Lett.} \textbf{\bibinfo{volume}{85}},
  \bibinfo{pages}{1791} (\bibinfo{year}{2000}).

\bibitem[{\citenamefont{Baron et~al.}(2012)\citenamefont{Baron, Campbell,
  DeMille, Doyle, Gabrielse, Gurevich, Hess, Hutzler, Kirilov, Kozyryev
  et~al.}}]{Baranov:2012}
\bibinfo{author}{\bibfnamefont{J.}~\bibnamefont{Baron}},
  \bibinfo{author}{\bibfnamefont{W.~C.} \bibnamefont{Campbell}},
  \bibinfo{author}{\bibfnamefont{D.}~\bibnamefont{DeMille}},
  \bibinfo{author}{\bibfnamefont{J.~M.} \bibnamefont{Doyle}},
  \bibinfo{author}{\bibfnamefont{G.}~\bibnamefont{Gabrielse}},
  \bibinfo{author}{\bibfnamefont{Y.~V.} \bibnamefont{Gurevich}},
  \bibinfo{author}{\bibfnamefont{P.~W.} \bibnamefont{Hess}},
  \bibinfo{author}{\bibfnamefont{N.~R.} \bibnamefont{Hutzler}},
  \bibinfo{author}{\bibfnamefont{E.}~\bibnamefont{Kirilov}},
  \bibinfo{author}{\bibfnamefont{I.}~\bibnamefont{Kozyryev}},
  \bibnamefont{et~al.}, \bibinfo{journal}{Chem. Rev.}
  \textbf{\bibinfo{volume}{112}}, \bibinfo{pages}{5012} (\bibinfo{year}{2012}).

\bibitem[{\citenamefont{DeMille}(2002)}]{DeMille:2002}
\bibinfo{author}{\bibfnamefont{D.}~\bibnamefont{DeMille}},
  \bibinfo{journal}{Phys. Rev. Lett.} \textbf{\bibinfo{volume}{88}},
  \bibinfo{pages}{067901} (\bibinfo{year}{2002}).

\bibitem[{\citenamefont{Yelin et~al.}(2006)\citenamefont{Yelin, Kirby, and
  Cot\'e}}]{Yelin:2006}
\bibinfo{author}{\bibfnamefont{S.~F.} \bibnamefont{Yelin}},
  \bibinfo{author}{\bibfnamefont{K.}~\bibnamefont{Kirby}}, \bibnamefont{and}
  \bibinfo{author}{\bibfnamefont{R.}~\bibnamefont{Cot\'e}},
  \bibinfo{journal}{Phys. Rev. A} \textbf{\bibinfo{volume}{74}},
  \bibinfo{pages}{050301(R)} (\bibinfo{year}{2006}).

\bibitem[{\citenamefont{Aldegunde et~al.}(2008)\citenamefont{Aldegunde,
  Rivington, \.Zuchowski, and Hutson}}]{Aldegunde:polar:2008}
\bibinfo{author}{\bibfnamefont{J.}~\bibnamefont{Aldegunde}},
  \bibinfo{author}{\bibfnamefont{B.~A.} \bibnamefont{Rivington}},
  \bibinfo{author}{\bibfnamefont{P.~S.} \bibnamefont{\.Zuchowski}},
  \bibnamefont{and} \bibinfo{author}{\bibfnamefont{J.~M.}
  \bibnamefont{Hutson}}, \bibinfo{journal}{Phys. Rev. A}
  \textbf{\bibinfo{volume}{78}}, \bibinfo{pages}{033434}
  (\bibinfo{year}{2008}).

\bibitem[{\citenamefont{Aldegunde et~al.}(2009)\citenamefont{Aldegunde, Ran,
  and Hutson}}]{Aldegunde:spectra:2009}
\bibinfo{author}{\bibfnamefont{J.}~\bibnamefont{Aldegunde}},
  \bibinfo{author}{\bibfnamefont{H.}~\bibnamefont{Ran}}, \bibnamefont{and}
  \bibinfo{author}{\bibfnamefont{J.~M.} \bibnamefont{Hutson}},
  \bibinfo{journal}{Phys. Rev. A} \textbf{\bibinfo{volume}{80}},
  \bibinfo{pages}{043410} (\bibinfo{year}{2009}).

\bibitem[{\citenamefont{Ran et~al.}(2010)\citenamefont{Ran, Aldegunde, and
  Hutson}}]{Ran:2010}
\bibinfo{author}{\bibfnamefont{H.}~\bibnamefont{Ran}},
  \bibinfo{author}{\bibfnamefont{J.}~\bibnamefont{Aldegunde}},
  \bibnamefont{and} \bibinfo{author}{\bibfnamefont{J.~M.}
  \bibnamefont{Hutson}}, \bibinfo{journal}{New J. Phys.}
  \textbf{\bibinfo{volume}{12}}, \bibinfo{pages}{043015}
  (\bibinfo{year}{2010}).

\bibitem[{\citenamefont{Gregory et~al.}(2017)\citenamefont{Gregory, Blackmore,
  Aldegunde, Hutson, and Cornish}}]{Gregory:RbCs-AC-Stark:2017}
\bibinfo{author}{\bibfnamefont{P.~D.} \bibnamefont{Gregory}},
  \bibinfo{author}{\bibfnamefont{J.~A.} \bibnamefont{Blackmore}},
  \bibinfo{author}{\bibfnamefont{J.}~\bibnamefont{Aldegunde}},
  \bibinfo{author}{\bibfnamefont{J.~M.} \bibnamefont{Hutson}},
  \bibnamefont{and} \bibinfo{author}{\bibfnamefont{S.~L.}
  \bibnamefont{Cornish}}, \bibinfo{journal}{Phys. Rev. A}
  \textbf{\bibinfo{volume}{96}}, \bibinfo{pages}{021402(R)}
  (\bibinfo{year}{2017}).

\bibitem[{\citenamefont{Will et~al.}(2016)\citenamefont{Will, Park, Yan, Loh,
  and Zwierlein}}]{Will:2016}
\bibinfo{author}{\bibfnamefont{S.~A.} \bibnamefont{Will}},
  \bibinfo{author}{\bibfnamefont{J.~W.} \bibnamefont{Park}},
  \bibinfo{author}{\bibfnamefont{Z.~Z.} \bibnamefont{Yan}},
  \bibinfo{author}{\bibfnamefont{H.}~\bibnamefont{Loh}}, \bibnamefont{and}
  \bibinfo{author}{\bibfnamefont{M.~W.} \bibnamefont{Zwierlein}},
  \bibinfo{journal}{Phys. Rev. Lett.} \textbf{\bibinfo{volume}{116}},
  \bibinfo{pages}{225306} (\bibinfo{year}{2016}).

\bibitem[{\citenamefont{Gregory et~al.}(2016)\citenamefont{Gregory, Aldegunde,
  Hutson, and Cornish}}]{Gregory:RbCs-microwave:2016}
\bibinfo{author}{\bibfnamefont{P.~D.} \bibnamefont{Gregory}},
  \bibinfo{author}{\bibfnamefont{J.}~\bibnamefont{Aldegunde}},
  \bibinfo{author}{\bibfnamefont{J.~M.} \bibnamefont{Hutson}},
  \bibnamefont{and} \bibinfo{author}{\bibfnamefont{S.~L.}
  \bibnamefont{Cornish}}, \bibinfo{journal}{Phys. Rev. A}
  \textbf{\bibinfo{volume}{94}}, \bibinfo{pages}{041403(R)}
  (\bibinfo{year}{2016}).

\bibitem[{\citenamefont{Ramsey}(1952)}]{Ramsey:1952}
\bibinfo{author}{\bibfnamefont{N.~F.} \bibnamefont{Ramsey}},
  \bibinfo{journal}{Phys. Rev.} \textbf{\bibinfo{volume}{85}},
  \bibinfo{pages}{60} (\bibinfo{year}{1952}).

\bibitem[{\citenamefont{Brown and Carrington}(2003)}]{Brown:2003}
\bibinfo{author}{\bibfnamefont{J.~M.} \bibnamefont{Brown}} \bibnamefont{and}
  \bibinfo{author}{\bibfnamefont{A.}~\bibnamefont{Carrington}},
  \emph{\bibinfo{title}{Rotational Spectroscopy of Diatomic Molecules}}
  (\bibinfo{publisher}{Cambridge University Press},
  \bibinfo{address}{Cambridge}, \bibinfo{year}{2003}).

\bibitem[{\citenamefont{Bryce and Wasylishen}(2003)}]{Bryce:2003}
\bibinfo{author}{\bibfnamefont{D.~L.} \bibnamefont{Bryce}} \bibnamefont{and}
  \bibinfo{author}{\bibfnamefont{R.~E.} \bibnamefont{Wasylishen}},
  \bibinfo{journal}{Acc. Chem. Res.} \textbf{\bibinfo{volume}{36}},
  \bibinfo{pages}{327} (\bibinfo{year}{2003}).

\bibitem[{\citenamefont{te~Velde et~al.}(2001)\citenamefont{te~Velde,
  Bickelhaupt, van Gisbergen, Fonseca~Guerra, Baerends, Snijders, and
  Ziegler}}]{ADF1}
\bibinfo{author}{\bibfnamefont{G.}~\bibnamefont{te~Velde}},
  \bibinfo{author}{\bibfnamefont{F.~M.} \bibnamefont{Bickelhaupt}},
  \bibinfo{author}{\bibfnamefont{S.~J.~A.} \bibnamefont{van Gisbergen}},
  \bibinfo{author}{\bibfnamefont{C.}~\bibnamefont{Fonseca~Guerra}},
  \bibinfo{author}{\bibfnamefont{E.~J.} \bibnamefont{Baerends}},
  \bibinfo{author}{\bibfnamefont{J.~G.} \bibnamefont{Snijders}},
  \bibnamefont{and} \bibinfo{author}{\bibfnamefont{T.}~\bibnamefont{Ziegler}},
  \bibinfo{journal}{J. Comput. Chem.} \textbf{\bibinfo{volume}{22}},
  \bibinfo{pages}{931} (\bibinfo{year}{2001}).

\bibitem[{ADF(2007)}]{ADF3}
\emph{\bibinfo{title}{{ADF2007.01}}},
  \bibinfo{howpublished}{http://www.scm.com} (\bibinfo{year}{2007}),
  \bibinfo{note}{{SCM}, {T}heoretical {C}hemistry{,} {V}rije {U}niversiteit{,}
  {A}msterdam{,} {T}he {N}etherlands}.

\bibitem[{\citenamefont{van Lenthe et~al.}(1993)\citenamefont{van Lenthe,
  Baerends, and Snijders}}]{vanLenthe:1993}
\bibinfo{author}{\bibfnamefont{E.}~\bibnamefont{van Lenthe}},
  \bibinfo{author}{\bibfnamefont{E.~J.} \bibnamefont{Baerends}},
  \bibnamefont{and} \bibinfo{author}{\bibfnamefont{J.~G.}
  \bibnamefont{Snijders}}, \bibinfo{journal}{J. Chem. Phys.}
  \textbf{\bibinfo{volume}{99}}, \bibinfo{pages}{4597} (\bibinfo{year}{1993}).

\bibitem[{\citenamefont{van Lenthe et~al.}(1994)\citenamefont{van Lenthe,
  Baerends, and Snijders}}]{vanLenthe:1994}
\bibinfo{author}{\bibfnamefont{E.}~\bibnamefont{van Lenthe}},
  \bibinfo{author}{\bibfnamefont{E.~J.} \bibnamefont{Baerends}},
  \bibnamefont{and} \bibinfo{author}{\bibfnamefont{J.~G.}
  \bibnamefont{Snijders}}, \bibinfo{journal}{J. Chem. Phys.}
  \textbf{\bibinfo{volume}{101}}, \bibinfo{pages}{9783} (\bibinfo{year}{1994}).

\bibitem[{\citenamefont{van Lenthe et~al.}(1999)\citenamefont{van Lenthe,
  Baerends, and Snijders}}]{vanLenthe:1999}
\bibinfo{author}{\bibfnamefont{E.}~\bibnamefont{van Lenthe}},
  \bibinfo{author}{\bibfnamefont{E.~J.} \bibnamefont{Baerends}},
  \bibnamefont{and} \bibinfo{author}{\bibfnamefont{J.~G.}
  \bibnamefont{Snijders}}, \bibinfo{journal}{J. Chem. Phys.}
  \textbf{\bibinfo{volume}{110}}, \bibinfo{pages}{8943} (\bibinfo{year}{1999}).

\bibitem[{\citenamefont{Lee et~al.}(1988)\citenamefont{Lee, Yang, and
  Parr}}]{lee:1988}
\bibinfo{author}{\bibfnamefont{C.}~\bibnamefont{Lee}},
  \bibinfo{author}{\bibfnamefont{W.}~\bibnamefont{Yang}}, \bibnamefont{and}
  \bibinfo{author}{\bibfnamefont{R.~G.} \bibnamefont{Parr}},
  \bibinfo{journal}{Phys. Rev. B} \textbf{\bibinfo{volume}{37}},
  \bibinfo{pages}{785} (\bibinfo{year}{1988}).

\bibitem[{\citenamefont{Becke}(1993)}]{becke:1993}
\bibinfo{author}{\bibfnamefont{A.~D.} \bibnamefont{Becke}},
  \bibinfo{journal}{J. Chem. Phys.} \textbf{\bibinfo{volume}{98}},
  \bibinfo{pages}{5648} (\bibinfo{year}{1993}).

\bibitem[{\citenamefont{Perdew et~al.}(1996)\citenamefont{Perdew, Burke, and
  Ernzerhof}}]{perdew:1996}
\bibinfo{author}{\bibfnamefont{J.~P.} \bibnamefont{Perdew}},
  \bibinfo{author}{\bibfnamefont{K.}~\bibnamefont{Burke}}, \bibnamefont{and}
  \bibinfo{author}{\bibfnamefont{M.}~\bibnamefont{Ernzerhof}},
  \bibinfo{journal}{Phys. Rev. Lett.} \textbf{\bibinfo{volume}{77}},
  \bibinfo{pages}{3865} (\bibinfo{year}{1996}).

\bibitem[{\citenamefont{Keal and Tozer}(2003)}]{Keal:2003}
\bibinfo{author}{\bibfnamefont{T.~W.} \bibnamefont{Keal}} \bibnamefont{and}
  \bibinfo{author}{\bibfnamefont{D.~J.} \bibnamefont{Tozer}},
  \bibinfo{journal}{J. Chem. Phys.} \textbf{\bibinfo{volume}{119}},
  \bibinfo{pages}{3015} (\bibinfo{year}{2003}).

\bibitem[{\citenamefont{Mills et~al.}(1988)\citenamefont{Mills, Cvita\v{s},
  Homann, Kallay, and Kuchitsu}}]{Mills:1988}
\bibinfo{author}{\bibfnamefont{I.}~\bibnamefont{Mills}},
  \bibinfo{author}{\bibfnamefont{T.}~\bibnamefont{Cvita\v{s}}},
  \bibinfo{author}{\bibfnamefont{K.}~\bibnamefont{Homann}},
  \bibinfo{author}{\bibfnamefont{N.}~\bibnamefont{Kallay}}, \bibnamefont{and}
  \bibinfo{author}{\bibfnamefont{K.}~\bibnamefont{Kuchitsu}},
  \emph{\bibinfo{title}{Quantities, Units and Symbols in Physical Chemistry}}
  (\bibinfo{publisher}{Blackwell}, \bibinfo{address}{Oxford},
  \bibinfo{year}{1988}).

\bibitem[{\citenamefont{Stone}(2016)}]{Stone:2016}
\bibinfo{author}{\bibfnamefont{N.~J.} \bibnamefont{Stone}},
  \bibinfo{journal}{At. Data Nucl. Data Tables}
  \textbf{\bibinfo{volume}{111-112}}, \bibinfo{pages}{1}
  (\bibinfo{year}{2016}).

\bibitem[{\citenamefont{Engelke et~al.}(1982)\citenamefont{Engelke, Ennen, and
  Meiwes}}]{Engelke:1982}
\bibinfo{author}{\bibfnamefont{F.}~\bibnamefont{Engelke}},
  \bibinfo{author}{\bibfnamefont{G.}~\bibnamefont{Ennen}}, \bibnamefont{and}
  \bibinfo{author}{\bibfnamefont{K.}~\bibnamefont{Meiwes}},
  \bibinfo{journal}{Chem. Phys.} \textbf{\bibinfo{volume}{66}},
  \bibinfo{pages}{391} (\bibinfo{year}{1982}).

\bibitem[{\citenamefont{Martin et~al.}(2001)\citenamefont{Martin, Crozet, Ross,
  Aubert-Frecon, Kowalczyk, Jastrzebski, and Pashov}}]{Martin:2001}
\bibinfo{author}{\bibfnamefont{F.}~\bibnamefont{Martin}},
  \bibinfo{author}{\bibfnamefont{P.}~\bibnamefont{Crozet}},
  \bibinfo{author}{\bibfnamefont{A.}~\bibnamefont{Ross}},
  \bibinfo{author}{\bibfnamefont{M.}~\bibnamefont{Aubert-Frecon}},
  \bibinfo{author}{\bibfnamefont{P.}~\bibnamefont{Kowalczyk}},
  \bibinfo{author}{\bibfnamefont{W.}~\bibnamefont{Jastrzebski}},
  \bibnamefont{and} \bibinfo{author}{\bibfnamefont{A.}~\bibnamefont{Pashov}},
  \bibinfo{journal}{J. Chem. Phys.} \textbf{\bibinfo{volume}{115}},
  \bibinfo{pages}{4118} (\bibinfo{year}{2001}).

\bibitem[{\citenamefont{Korek et~al.}(2000)\citenamefont{Korek, Allouche,
  Kobeissi, Chaalan, Dagher, Fakherddin, and Aubert-Frecon}}]{Korek:2000}
\bibinfo{author}{\bibfnamefont{M.}~\bibnamefont{Korek}},
  \bibinfo{author}{\bibfnamefont{A.~R.} \bibnamefont{Allouche}},
  \bibinfo{author}{\bibfnamefont{M.}~\bibnamefont{Kobeissi}},
  \bibinfo{author}{\bibfnamefont{A.}~\bibnamefont{Chaalan}},
  \bibinfo{author}{\bibfnamefont{M.}~\bibnamefont{Dagher}},
  \bibinfo{author}{\bibfnamefont{K.}~\bibnamefont{Fakherddin}},
  \bibnamefont{and}
  \bibinfo{author}{\bibfnamefont{M.}~\bibnamefont{Aubert-Frecon}},
  \bibinfo{journal}{Chem. Phys.} \textbf{\bibinfo{volume}{256}},
  \bibinfo{pages}{1} (\bibinfo{year}{2000}).

\bibitem[{\citenamefont{Staanum et~al.}(2007)\citenamefont{Staanum, Pashov,
  Kn\"ockel, and Tiemann}}]{Staanum:2007}
\bibinfo{author}{\bibfnamefont{P.}~\bibnamefont{Staanum}},
  \bibinfo{author}{\bibfnamefont{A.}~\bibnamefont{Pashov}},
  \bibinfo{author}{\bibfnamefont{H.}~\bibnamefont{Kn\"ockel}},
  \bibnamefont{and} \bibinfo{author}{\bibfnamefont{E.}~\bibnamefont{Tiemann}},
  \bibinfo{journal}{Phys. Rev. A} \textbf{\bibinfo{volume}{75}},
  \bibinfo{pages}{042513} (\bibinfo{year}{2007}).

\bibitem[{\citenamefont{Wormsbecher et~al.}(1981)\citenamefont{Wormsbecher,
  Hessel, and Lovas}}]{Wormsbecher:1981}
\bibinfo{author}{\bibfnamefont{R.}~\bibnamefont{Wormsbecher}},
  \bibinfo{author}{\bibfnamefont{M.}~\bibnamefont{Hessel}}, \bibnamefont{and}
  \bibinfo{author}{\bibfnamefont{F.}~\bibnamefont{Lovas}}, \bibinfo{journal}{J.
  Chem. Phys.} \textbf{\bibinfo{volume}{74}}, \bibinfo{pages}{6983}
  (\bibinfo{year}{1981}).

\bibitem[{\citenamefont{Docenko et~al.}(2004)\citenamefont{Docenko, Tamanis,
  Ferber, Pashov, Kn\"ockel, and Tiemann}}]{Docenko:NaRb:2004}
\bibinfo{author}{\bibfnamefont{O.}~\bibnamefont{Docenko}},
  \bibinfo{author}{\bibfnamefont{M.}~\bibnamefont{Tamanis}},
  \bibinfo{author}{\bibfnamefont{R.}~\bibnamefont{Ferber}},
  \bibinfo{author}{\bibfnamefont{A.}~\bibnamefont{Pashov}},
  \bibinfo{author}{\bibfnamefont{H.}~\bibnamefont{Kn\"ockel}},
  \bibnamefont{and} \bibinfo{author}{\bibfnamefont{E.}~\bibnamefont{Tiemann}},
  \bibinfo{journal}{Phys. Rev. A} \textbf{\bibinfo{volume}{69}},
  \bibinfo{pages}{042503} (\bibinfo{year}{2004}).

\bibitem[{\citenamefont{Docenko et~al.}(2006)\citenamefont{Docenko, Tamanis,
  Zaharova, Ferber, Pashov, Kn\"ockel, and Tiemann}}]{Docenko:2006}
\bibinfo{author}{\bibfnamefont{O.}~\bibnamefont{Docenko}},
  \bibinfo{author}{\bibfnamefont{M.}~\bibnamefont{Tamanis}},
  \bibinfo{author}{\bibfnamefont{J.}~\bibnamefont{Zaharova}},
  \bibinfo{author}{\bibfnamefont{R.}~\bibnamefont{Ferber}},
  \bibinfo{author}{\bibfnamefont{A.}~\bibnamefont{Pashov}},
  \bibinfo{author}{\bibfnamefont{H.}~\bibnamefont{Kn\"ockel}},
  \bibnamefont{and} \bibinfo{author}{\bibfnamefont{E.}~\bibnamefont{Tiemann}},
  \bibinfo{journal}{J. Phys. B} \textbf{\bibinfo{volume}{39}},
  \bibinfo{pages}{S929} (\bibinfo{year}{2006}).

\bibitem[{\citenamefont{Ross et~al.}(1990)\citenamefont{Ross, Effantin, Crozet,
  and Boursey}}]{Ross:1990}
\bibinfo{author}{\bibfnamefont{A.~J.} \bibnamefont{Ross}},
  \bibinfo{author}{\bibfnamefont{C.}~\bibnamefont{Effantin}},
  \bibinfo{author}{\bibfnamefont{P.}~\bibnamefont{Crozet}}, \bibnamefont{and}
  \bibinfo{author}{\bibfnamefont{E.}~\bibnamefont{Boursey}},
  \bibinfo{journal}{J. Phys. B} \textbf{\bibinfo{volume}{23}},
  \bibinfo{pages}{L247} (\bibinfo{year}{1990}).

\bibitem[{\citenamefont{Ferber et~al.}(2008)\citenamefont{Ferber, Klincare,
  Nikolayeva, Tamanis, Kn\"ockel, Tiemann, and Pashov}}]{Ferber:2008}
\bibinfo{author}{\bibfnamefont{R.}~\bibnamefont{Ferber}},
  \bibinfo{author}{\bibfnamefont{I.}~\bibnamefont{Klincare}},
  \bibinfo{author}{\bibfnamefont{O.}~\bibnamefont{Nikolayeva}},
  \bibinfo{author}{\bibfnamefont{M.}~\bibnamefont{Tamanis}},
  \bibinfo{author}{\bibfnamefont{H.}~\bibnamefont{Kn\"ockel}},
  \bibinfo{author}{\bibfnamefont{E.}~\bibnamefont{Tiemann}}, \bibnamefont{and}
  \bibinfo{author}{\bibfnamefont{A.}~\bibnamefont{Pashov}},
  \bibinfo{journal}{J. Chem. Phys.} \textbf{\bibinfo{volume}{128}},
  \bibinfo{pages}{244316} (\bibinfo{year}{2008}).

\bibitem[{\citenamefont{Kat\^o and Kobayashi}(1983)}]{Kato:1983}
\bibinfo{author}{\bibfnamefont{H.}~\bibnamefont{Kat\^o}} \bibnamefont{and}
  \bibinfo{author}{\bibfnamefont{H.}~\bibnamefont{Kobayashi}},
  \bibinfo{journal}{J. Chem. Phys.} \textbf{\bibinfo{volume}{79}},
  \bibinfo{pages}{123} (\bibinfo{year}{1983}).

\bibitem[{\citenamefont{Fedorov et~al.}(2014)\citenamefont{Fedorov, Derevianko,
  and Varganov}}]{Fedorov:2014}
\bibinfo{author}{\bibfnamefont{D.~A.} \bibnamefont{Fedorov}},
  \bibinfo{author}{\bibfnamefont{A.}~\bibnamefont{Derevianko}},
  \bibnamefont{and} \bibinfo{author}{\bibfnamefont{S.~A.}
  \bibnamefont{Varganov}}, \bibinfo{journal}{J. Chem. Phys.}
  \textbf{\bibinfo{volume}{140}}, \bibinfo{pages}{184315}
  (\bibinfo{year}{2014}).

\end{thebibliography}

\end{document}